\documentstyle[12pt]{article}
\newcommand{\be}{\begin{equation}}
\newcommand{\ee}{\end{equation}}

\begin{document}

\centerline {\large {\bf CHAOTICITY AND DISSIPATION OF }}
\centerline {\large {\bf NUCLEAR COLLECTIVE MOTION}}
\centerline {\large {\bf IN A CLASSICAL MODEL}}

\bigskip
\bigskip

\centerline {M.\, Baldo, {\underline {G.\,F. Burgio}}, A.\, Rapisarda}

\centerline {\it Istituto Nazionale di Fisica Nucleare, Sezione di Catania}

\centerline {\it and Dipartimento di Fisica, Universit\'a di Catania }

\centerline {\it  Corso Italia 57, I-95129 Catania, Italy } 

\bigskip

\centerline {and P.\, Schuck }

\centerline {\it Institut de Physique Nucl\'eaire, Universit\'e de Grenoble}

\centerline {\it 53 Avenue des Martyrs, 38026 Grenoble Cedex, France}

\bigskip
\bigskip

\begin{abstract} 
We analyze the behavior of a gas of classical particles
moving in a two-dimensional "nuclear" billiard 
whose multipole-deformed walls undergo periodic shape oscillations.
We demonstrate that a single particle Hamiltonian containing
coupling terms between the particles' motion and the collective coordinate 
induces a chaotic dynamics for any multipolarity, independently
on the geometry of the billiard. The absence of coupling terms allows us
to recover qualitatively the "wall formula" predictions.
We also discuss the dissipative 
behavior of the wall motion and its relation with the order-to-chaos
transition in the dynamics of the microscopic degrees of freedom.
\end{abstract}

\bigskip
\bigskip

\section{Introduction}

In the last twenty years the dissipation of collective motion in nuclei has 
been widely observed \cite{sp81} in low energy particle and heavy ion
collisions and it still represents a theoretically 
unsolved problem.  It is commonly believed that both one-body processes, 
{\it i.e.} collisions of nucleons with the nuclear wall generated by the common
selfconsistent mean field, and two-body collisions produce
dissipation, although their interplay is not well known.

In this work we discuss only one-body processes. In this framework,
Blocki {\it et al.}\cite{bswi} analyzed the behavior of a gas
of classical non-interacting particles enclosed in a 
multipole-deformed container which undergoes periodic shape 
oscillations. Particles move on linear trajectories and collide 
elastically against the walls. In this model wall and particles' motion is
uncoupled and therefore the wall oscillates at the same frequency
pumping energy into the gas. 
For this system, the authors study the increase of the particles' 
kinetic energy as a function of time.
They find that for the quadrupole case the gas does not heat up, 
whereas for higher multipolarities the gas kinetic energy increases 
with time, in agreement with the "wall formula" predictions \cite{wf}. 
They attribute the different
behavior to the fact that for low deformations the particles' motion
is regular and corresponds to an integrable situation, whereas for 
higher multipolarities the strong shape irregularities
leads to divergence between trajectories and therefore
to chaotic motion.
Although their results look very interesting,
their application to the nuclear case is not straightforward because 
i) the selfconsistent mean field is absent, ii) the total energy 
is not conserved.\par
A step forward in this direction has been performed by Bauer
{\it et al.} in ref.\cite{bauer}. In this work the authors study
the damping of collective motion in nuclei within the 
semiclassical Vlasov equation. Here
selfconsistency is taken into account and the total energy is 
conserved. A multipole-multipole interaction of the Bohr-Mottelson
type is adopted for quadrupole and octupole deformations.
In both cases the dynamical evolution shows a regular undamped 
collective motion which coexists with a weakly chaotic single-particle 
dynamics.

In order to clarify the relationship between chaos at the microscopic
level and damping of the collective motion, we consider 
a gas of classical non-interacting particles moving in a two-dimensional
billiard. Particles collide with the oscillating
walls and transfer energy to it, therefore collisions are always 
inelastic. We consider the gas + billiard as an Hamiltonian system, therefore 
the total energy conservation imposes that the walls can give back 
energy to the gas. Therefore the main feature of the model is that 
particles and walls are strongly coupled, and this has far reaching
consequences. The most important one is a) chaos shows up in the 
single particle motion for any surface deformation and b) the 
dissipation of the collective motion does not depend on the geometry
of the billiard. We describe our model in section 2,
and the results in section 3. In section 4 we draw some conclusions.

\bigskip
\bigskip

\section {The Model}

In ref.\cite{bbrs} we considered a classical
version of the vibrating potential model for finite 
nuclei (see {\it e.g.} ref.\cite{risc}). In this model several 
non-interacting classical particles move 
in a two-dimensional deep potential well and hit the oscillating 
surface. Using polar coordinates, the Hamiltonian depends on a set of
$\{r_i, \theta_i \}$ variables, describing
the motion of the particles, and the collective coordinate $\alpha$. 
The Hamiltonian reads

\be
\label {1}
H(r_i, \theta_i, \alpha) = \sum_{i=1}^A ({{p_{r_i}^2}\over {2m}} +  
{{p_{\theta_i}^2} \over{2mr_i^2}} +
V(r_i, R(\theta_i))) + {{p_\alpha^2} \over{2M}} + {1\over 2}
M \Omega^2 \alpha^2
\ee

\noindent
being $\{p_{r_i}, p_{\theta_i}, p_{\alpha}\}$ the conjugate momenta of 
$\{r_i, \theta_i, \alpha\}$.
$m = 938~MeV$ is the nucleon mass, and $M = \eta m A R_o^2$ is the Inglis 
mass, chosen proportional to the total number
of particles $A$ and to the circular billiard radius $R_o$.
The factor $\eta$ ensures that during the dynamical evolution the equilibrium 
fluctuations are small. In our case $\eta = 1$ for the monopole oscillation,
whereas $\eta = 10$ for quadrupole and octupole. Therefore in the
$L=0$ case, collisions of particles against the walls are more inelastic.
\noindent
$\Omega$ is the oscillation frequency of the collective variable 
$\alpha$. The potential $V(r, R(\theta))$ is zero inside the billiard 
and a very steeply rising function on the surface, 
${V(r, R(\theta))} = {V_o \over
{(1 + exp({{R(\theta) - r} \over {a}}))}}$, with $V_o = 1500~MeV$ and 
$a = 0.01~fm$.
The surface is described by
$R(\theta) = R_o (1 + \alpha P_L(cos\theta))$, where $P_L$ is 
the Legendre polynomial with multipolarity $L$.
Therefore this potential couples the collective variable motion to 
the particles' dynamics and prevents particles from escaping. 
The numerical simulation is based on the
the Hamilton's equations, which are solved 
with an algorithm of fourth-order 
Runge-Kutta type with typical time step sizes of $1~fm/c$.
All calculations were performed with a number
of particles $A = 30$. The total energy was conserved with high accuracy.

As far as the initial conditions are concerning, we assign random
positions to the particles inside the billiard and random initial
momenta according to a two-dimensional Maxwell-Boltzmann distribution 
with a temperature  $T = 36~MeV$. 
We consider the wall oscillation taking place close to adiabatic 
conditions. For this purpose we impose a wall frequency smaller than 
the single particle one and choose $\Omega = 0.05~c/fm$,
which corresponds to a oscillation period 
$\tau_w = {2 \pi \over \Omega} \sim
125.66~fm/c$. The single particle oscillation period is equal to 
$\tau_p = {2 R_o \over v}$,
being $v$ the most probable particle speed, $v = \sqrt {2T/m}$.
We choose $R_o = 6~fm$ and this
gives a single particle period $\tau_p \sim 43.3~fm/c$.

Since in the realistic nuclear case the collective motion takes place 
around equilibrium, 
the initial wall coordinate has been chosen equal to 
$\alpha_0 = \bar\alpha + \delta \alpha$, being $\bar \alpha$ 
the equilibrium value and $\delta \alpha$ a small deviation.
The equilibrium value $\bar \alpha$ corresponds of course to the
thermodinamic limit, which is actually reached when 
considering an ensemble of copies of the system,
all of them with an initial value $\alpha = \bar \alpha$ 
and differing from each other in the initial microscopic distribution
of particles' positions and momenta. It can be shown that the
equilibrium value actually depends on the considered multipolarity
\cite{bbrs_new}. Moreover at time $t=0$ we put $p_\alpha = 0$, the wall 
having only potential energy. After we checked that the numerical
simulation produces the good equilibrium properties, we slightly
perturbed the equilibrium collective coordinate $\bar \alpha$ 
by an amount $\delta \alpha = 0.3$, and let both the billiard and the
particles evolve in time.

\bigskip
\bigskip

\section {Results}

\subsection {Scatter plots}

One possible way in order to investigate the role played by coupling 
and see whether it can induce a chaotic dynamics, is drawing
Poincare's surface of sections for the single particle coordinate,
which is impossible to perform in our case because of the large 
number of degrees of freedom. An alternative way to visualize 
a chaotic behavior is to draw scatter plots, see Fig.1.
There we display the final radial coordinate at a time $t$ of one 
chosen particle vs. the one at $t=0$ for the monopole, quadrupole 
and octupole deformations at times $t=50, 200, 300 $ and $500 fm/c$.
These plots are very useful and are commonly used in transient irregular
situations like chaotic scattering \cite{scatt} and nuclear multifragmentation 
\cite{frag}. The idea 
is that if the dynamics is regular, two initially close points 
in space stay close even at later times, but if the dynamics is 
chaotic the two points will soon
separate due to the exponential divergence induced by chaos. 
In the first case this plot will show a regular curve,
whereas in the other one a diffused pattern appears.
We note that for all multipolarities the initially
regular curves change into scatter plots, which clearly show that 
the coupling to the wall oscillation randomizes the single particle
motion. This is at variance with what was discussed in ref.\cite{bswi}, 
where chaos is supposed to appear only for multipolarities $L > 2$. 
In our model the coupling between wall and particles' motion produces a 
chaotic dynamics even for $L \leq 2$. 
In addition, the higher the multipolarity the earlier chaos starts 
because of the increased shape irregularity.

\vskip 11cm
\noindent
Fig.1. {\footnotesize{The final radial coordinate for one particle is
drawn as a function of the initial one at different times t = 50, 200, 300,
500 fm/c. Calculations are performed for multipolarities L = 0, 2, 3.}}
\vskip 0.5cm

For this purpose, a more
quantitative analysis can be performed because scatter plots of Fig.1 
have a typical fractal structure. As already done in ref.\cite{frag},
a fractal correlation dimension $D_2$ can be calculated from the
correlation integral $C(r)$ \cite{grass}. One first counts how many points 
have a smaller distance than some given distance $r$. As $r$ varies, so does
$C(r)$, defined as

\be
\label{2}
C(r)~ = ~ {1\over {M^2}} \sum_{i,j}^M \Theta(r - |{\bf z_i} - {\bf z_j}|)
\ee
\noindent
being $\Theta$ the Heaviside step function and ${\bf z_i}$ a vector
whose two components ($x_i, y_i$) are the points coordinates.
M is the total number of points.
The fractal correlation dimension $D_2$ is then defined by

\be
\label{3}
D_2~ = ~\lim_{r\to~0}~ {log~C(r) \over {log~r}}
\ee

In Fig.2 we display $D_2$ vs. time for each multipolarity. At very beginning 
$D_2$ is equal to one, showing that the motion is regular. 
As time goes on, regularity is lost and the motion becomes chaotic 
until a complete randomness is reached, in which case $D_2~=~2$ as expected
\cite{frag}. This result confirms the one published in ref.\cite{bbrs},
where a different method of calculation was however employed.

\vskip 10cm
\noindent
Fig.2 : {\footnotesize{The time evolution of the fractal correlation 
dimension $D_2$ is plotted for the multipolarities L = 0, 2, 3.
The dashed lines are to guide the eye.}}
\vskip 0.5cm

\subsection {Chaos and dissipation}

Now let us discuss the dissipative properties of our system.
On the left-hand side of Fig.3 we plot the evolution of the collective 
variable $\alpha$ vs. time for one event only. Each panel
corresponds to a fixed multipolarity. We note that 
the amplitude of the collective motion shows an irregular oscillation,
at variance with the results found in ref.\cite{bauer}.
A slight damping can be observed, at variance with the results 
published in ref.\cite{bbrs} where calculations with a smaller number of
particles were performed.
Please note that $\alpha$ keeps on oscillating around the equilibrium
value $\bar \alpha$ which is equal to zero for $L=2, 3$, whereas
in the monopole case it differs from zero \cite{bbrs} and depends
on the gas temperature and on the wall frequency. 
On the right-hand side of Fig.3 we plot the time evolution of the excitation
energy of the gas, defined as the relative variation of the total energy
of the gas $E$ with respect to its initial value $E_0$. In all three cases 
the gas has been heated up, although
a monotonically increasing trend shows up for the quadrupole and octupole 
modes, while an irregular oscillating pattern is visible for the monopole case.
Moreover, in the $L=0$ oscillation the energy gained by the gas is lower
than in the $L=2, 3$ cases. Therefore some dissipation is present 
for all multipolarities and is larger for increased $L$.

\vskip 11cm
\noindent
Fig.3 : {\footnotesize{ On the left-hand side the time evolution of the
collective variable is shown for the multipolarities L = 0, 2, 3.
On the right-hand side the time evolution of the
gas excitation energy is shown for the same L.}}
\vskip 0.5cm

In order to have a good global picture 
of the macroscopic system, many events are needed 
and average observables should be considered. In Fig.4
we display the behavior of an ensemble of 1000 different events, 
each obtained by assigning random initial conditions to the particles 
both in coordinate and momentum space. Average quantities are reported.

The collective variable $\alpha$, plotted on the left-hand side of
Fig.4, shows a damping much stronger than the one
observed in the single event. This is due to the fact that,
once chaos has developed, the particles hit the wall essentially at random, 
and this produces a stochastic dephasing of the collective motion between 
different events. The overlap of many of those events produces then a 
strong damping. We note that the motion of the collective variable is 
completely damped out for times which depend on the multipolarity and 
develops around equilibrium. The reader should keep in mind that 
for $L=0$ collisions are more inelastic (see paragraph 2) due to the
lighter mass of the wall, therefore
the monopole oscillation damps out earlier than the $L=2$ mode.
The same mass was used in the calculations published in ref.\cite{bbrs}.
If we put $\eta = 10$ even for $L=0$, the damping time is longer
\cite{bbrs_new}.

\vskip 11cm
\noindent
Fig.4 : {\footnotesize{Same as Fig.3, but for an ensemble of 1000 events.}}
\vskip 0.5cm

As far as the excitation energy of the gas is concerned,
in the right-hand side of Fig.4 we note that the monopole oscillation 
behaves quite
differently than the ones with $L=2, 3$. In fact, while in those modes
the average among different events only decreases the fluctuations 
with respect to the single event, in $L=0$ the average
produces a dramatically different trend than in the single event. 
One could deduce that the stochastic dephasing plays a major role 
for the monopole oscillation. Moreover for $L=2, 3$ we note 
the presence of two different regimes: a first one lasting
for about 250 fm/c characterized by a sharp rising
in the excitation energy, and a second one at successive times 
where saturation dominates. As it can be deduced from Fig.1, the 
first stage is related to the onset of chaos in the single-particle motion. 
After that chaos has fully developed, and a large part of the total energy
has been pumped into the gas, a certain degree of
thermalization is reached and saturation shows up. This behavior 
is completely absent in the monopole case. Therefore it seems 
that different kinds of dissipation can originate from the same 
underlying chaotic single-particle motion.
It should be stressed that, within the time of chaos development, 
the wall has dissipated only a fraction of its energy, as it is 
clearly shown in Fig.5.

\vskip 10cm
\noindent
Fig.5 : {\footnotesize{ The wall energy, averaged over an ensemble of 
1000 events, is reported vs. time for different multipolarities.}}
\vskip 0.5cm

A completely different dynamics appears if the coupling terms in the
single-particle Hamiltonian (1) are neglected, thus allowing the wall
to pump energy into the particles' gas. On the contrary, the particles
cannot transfer energy to the wall, and therefore the whole
system is dissipative. In this case the
collective variable $\alpha$ is a regularly oscillating curve\cite{bbrs}.
In Fig.6 we show the time evolution of the excitation energy 
of the gas. While in the monopole case the gas has been slightly heated up,
a strong dissipation characterized by only one regime 
shows up for $L=2, 3$, thus recovering
qualitatively the trend predicted by the "wall formula" and examined in 
ref.\cite{wf}. 
This is also confirmed by the blow-up shown in Fig.7,
where we compare the different trends with coupling terms 
(left-hand side) and without (right-hand side) for quadrupole and octupole
modes. Dashed lines are best-fit curves. In the case with coupling we note 
that the dissipation rates related to the slopes $b$
are identical, while they strongly depend on the 
multipolarity in the uncoupled case.
This means that when including coupling terms the dynamics plays a major role,
independently on the surface irregularity, 
while in the uncoupled case the geometry of the problem is more important.

\vskip 7.5cm
\noindent
Fig.6 : {\footnotesize{ The excitation energy calculated without 
coupling terms is reported vs. time for different multipolarities.}}

\vskip 10.5cm
\noindent
Fig.7 : {\footnotesize{ A blow-up of fig.6 during the first 500 fm/c
is  displayed vs. time for L=2, 3. On the left-hand side results for
the coupled case are shown, while on the right-hand side coupling terms
are off. $b$ is the slope of the best-fit curves (dashed lines).}}
\vskip 1cm

\section {Conclusions}

In conclusions, we have presented a novel approach based on the
solution of the Hamilton's equations for several classical particles
moving in a classical billiard having nuclear-like dimensions, 
in order to explain dissipation of the collective 
motion. We found that the presence of 
a coupling term in the single particle Hamiltonian
induces chaotic motion at microscopic level. 

As far as the monopole mode is concerned, we found 
irregular behavior and a slight damping in the single event.
On the other hand, a whole bunch displays dissipation because of 
incoherence among different events. 
This incoherence is produced by the chaotic single particle dynamics, 
which makes all events belonging to the same ensemble strongly 
different one from each other. 
The dissipative process looks different for the quadrupole and octupole
modes. In fact, while the single event properties are similar to the the
monopole case, an ensemble of events shows that two different
regimes appear : a) an initial fast dissipative evolution 
corresponding to the onset of chaos in the single-particle motion
and b) a slower dissipative trend towards equilibrium.

A strongly different dynamics appears if the coupling terms are 
switched off. The excitation energy of the gas monotonically increases and 
only one dissipative regime shows up. 
However, the main result of our work is that dissipation is present
for {\it any} multipolarity, at variance with the prediction of
ref.\cite{bswi}.

One should realize that our model as well as the one of Swiatecki {\it et al.}
imply real particle wall collisions. Such particle wall collisions are absent 
in mean field calculations \cite{bauer}. It is unclear at this moment whether
this difference can have strong consequences on the damping mechanism 
of collectivity but this possibility is certainly not excluded. 
Corresponding studies are planned for the future.

We hope that the results obtained in such a schematic classical model 
could be of help in understanding the damping of nuclear collective motion.

\end{document}